\documentclass[12pt]{article}
\usepackage[english]{babel}
\usepackage[a4paper,top=2cm,bottom=2cm,left=3cm,right=3cm,marginparwidth=1.75cm]{geometry}
\usepackage[T1]{fontenc}
\usepackage{listings}

\usepackage{amsmath}
\usepackage{graphicx}
\usepackage[colorlinks=true, allcolors=blue]{hyperref}
\usepackage{authblk}
\usepackage{enumitem}
\usepackage[normalem]{ulem}

\newbox{\myorcidaffilbox}
\sbox{\myorcidaffilbox}{\large\includegraphics[height=1.7ex]{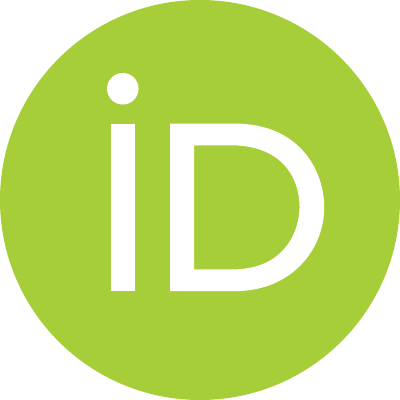}}
\newcommand{\orcidaffil}[1]{%
  \href{https://orcid.org/#1}{\usebox{\myorcidaffilbox}}}

\usepackage[
backend=biber,
sorting=none,
style=nature
]{biblatex}
\usepackage{amssymb}% http://ctan.org/pkg/amssymb
\usepackage{pifont}% http://ctan.org/pkg/pifont
\newcommand{\cmark}{\ding{51}}%
\newcommand{\xmark}{\ding{55}}%

\newcommand\update[1]{#1}

\title{LAPIS is a fast web API for massive open virus sequencing data}

\author[1,2,$\dagger$,a]{\orcidaffil{0000-0002-8763-2937}~Chaoran~Chen}
\author[1,3]{\orcidaffil{0000-0002-2177-7092}~Alexander~Taepper}
\author[4]{\orcidaffil{0000-0003-2693-084X}~Fabian~Engelniederhammer}
\author[4]{\orcidaffil{0000-0001-8033-2811}~Jonas~Kellerer}
\author[5,2]{\orcidaffil{0000-0002-6138-6539}~Cornelius~Roemer}
\author[1,2,$\dagger$,b]{\orcidaffil{0000-0001-6431-535X}~Tanja~Stadler}
\affil[1]{Department of Biosystems Science and Engineering, ETH Zürich, Basel, Switzerland}
\affil[2]{Swiss Institute of Bioinformatics, Basel, Switzerland}
\affil[3]{School of Computation, Information and Technology - Informatics, TU Munich, Munich, Germany}
\affil[4]{TNG Technology Consulting GmbH, Unterföhring, Germany}
\affil[5]{Biozentrum, University of Basel, Basel, Switzerland}
\affil[$\dagger$]{Corresponding author}
\affil[a]{chaoran.chen@bsse.ethz.ch}
\affil[b]{tanja.stadler@bsse.ethz.ch}

\usepackage{lineno}

\begin{document}
\maketitle

\begin{abstract}

\textbf{Background:} Recent epidemic outbreaks such as the SARS-CoV-2 pandemic and the mpox outbreak in 2022 have demonstrated the value of genomic sequencing data for tracking the origin and spread of pathogens. Laboratories around the globe generated new sequences at unprecedented speed and volume and bioinformaticians developed new tools and dashboards to analyze this wealth of data. However, a major challenge that remains is the lack of simple and efficient approaches for accessing and processing sequencing data.

\textbf{Results:} The Lightweight API for Sequences (LAPIS) facilitates rapid retrieval and analysis of genomic sequencing data through a REST API. It supports complex mutation- and metadata-based queries and can perform aggregation operations on massive datasets. LAPIS is optimized for typical questions relevant to genomic epidemiology. Using a newly-developed in-memory database engine, it has a high speed and throughput: between 25~January and 4~February 2023, the SARS-CoV-2 instance of LAPIS, which contains 14.5 million sequences, processed over 20~million requests with a mean response time of 411~ms and a median response time of 1~ms. LAPIS is the core engine behind our dashboards on genspectrum.org and we currently maintain public LAPIS instances for SARS-CoV-2 and mpox.

\textbf{Conclusions:} Powered by an optimized database engine and available through a web API, LAPIS enhances the accessibility of genomic sequencing data. It is designed to serve as a common backend for dashboards and analyses with the potential to be integrated into common database platforms such as GenBank.

\end{abstract}

\section{Background}

% Public health importance
Pathogen genomic sequencing data are a key public health resource for responding to epidemic outbreaks. During the early stages of an outbreak, genomic sequencing data are essential for understanding the origin, evolution, and extent of spread of the pathogen \cite{Li_2021, Knyazev2022}. At later stages, sequencing data are the primary early indicator of evolutionary and epidemiological changes, as demonstrated repeatedly with SARS-CoV-2 variants \cite{Viana2022, Chen2021-b117, Tegally2022}. Rapid analysis of sequencing data is therefore a crucial component for evidence-based public health responses. Although a lot of infrastructure for generating and analyzing genomic sequencing data in real-time was established during the SARS-CoV-2 pandemic, major challenges remain \cite{Black2020, Hodcroft2021, Knyazev2022, Chen2022-challenges}.

% Dashboards and notebooks are core tools for analyzing data
The unprecedented scale of SARS-CoV-2 sequence generation, coupled with enormous popular interest in these data, highlights a need for user-friendly tools for analyzing massive sequence data sets. One such category of tools is web dashboards. Once set up, these can be used by a wide audience without requiring programming and data science knowledge. Examples of popular dashboards that digest massive SARS-CoV-2 data sets include the CDC's COVID Data Tracker \cite{cdc-tracker}, CoVariants \cite{Hodcroft_CoVariants2021}, Outbreak.info \cite{Gangavarapu-outbreakinfo}, and our own CoV-Spectrum dashboard \cite{Chen2021-covspectrum}. Another category of tools that facilitate quick, ad-hoc analyses are ``notebooks'' like Jupyter Notebooks and R Markdown scripts. Notebooks are useful to data scientists with programming knowledge to quickly perform their own statistical analyses and generate their own plots. Combined, dashboards and notebooks allow different users to access different visualizations and focus on different aspects of the data. In this way, everyone from experts like scientists and public health agencies to the general public can benefit from sequence data.

% ...and they have similar needs/perform similar basic operations - and the challenge caused by the data size
Many of these tools for sequence data analysis require common operations on sequence data like filtering, stratification, and aggregation. For instance, filtering for sequences with certain mutations and calculating the relative frequency of mutations are commonly performed operations for genome sequencing data. Although these operations are simple in principle, the gigantic size of modern genome sequence data sets makes them non-trivial. Over 14 million SARS-CoV-2 sequences are available, and up to hundreds of thousands of new sequences are added weekly. General-purpose database systems such as PostgreSQL are not optimized for genomic sequence analysis on this scale.

% Brief Summary of LAPIS
Our resource LAPIS (Lightweight API for Sequences) is designed to perform common data operations on millions of genomic sequences within milliseconds, facilitating interactive data exploration. Using a self-written in-memory database engine, it is optimized for filtering and aggregating large genomic sequencing data sets. Accessible through a web API (application programming interface), we believe that LAPIS can serve as a common backend for many dashboards and analyses (e.g., through notebooks). This would relieve scientists and dashboard builders from the costly but boring task of developing their own databases and implementing common basic operations. Instead, they would be free to focus on analysis and visualization tasks. Furthermore, LAPIS streamlines the direct download of cleaned and pre-processed data including aligned and unaligned sequences.

% Data service, not a database
In contrast to data repositories like GenBank \cite{Benson2017}, LAPIS is not a broad database but a targeted data service. While GenBank contains sequences from more than 400,000 species and aims to provide a general and stable data source, LAPIS  supports features specific to an outbreak species like lineage/clade annotation and filtering by mutations from a reference genome. In this way, LAPIS aims to support answering current research and public health questions about emerging pathogen threats.

\section{Results}

\subsection{Functionalities}

LAPIS implements many of the same functionalities as GenBank and additionally supports novel download, filter, and aggregation functionalities to support outbreak analysis (Table \ref{tab:comparison_genbank}).

\begin{table}[h]
\caption{Feature comparisons between GenBank and LAPIS.}
\centering
\begin{tabular}{|l|c|c|}
\hline
Feature                                        & GenBank & LAPIS  \\ \hline
Download metadata                              & \cmark  & \cmark \\ \hline
Download unaligned sequences                   & \cmark  & \cmark \\ \hline
Download aligned sequences                     & \xmark  & \cmark \\ \hline
Download protein amino acid sequences          & \xmark  & \cmark \\ \hline
Download mutations                             & \xmark  & \cmark \\ \hline
Filter by basic metadata (country, date, etc.) & \cmark  & \cmark \\ \hline
Filter by lineage/clade                        & \cmark  & \cmark \\ \hline
Filter by mutations                            & \xmark  & \cmark \\ \hline
Perform aggregation                            & \xmark  & \cmark \\ \hline
\end{tabular}
\label{tab:comparison_genbank}
\end{table}

\begin{figure}[h]
    \centering
    \includegraphics[width=\textwidth]{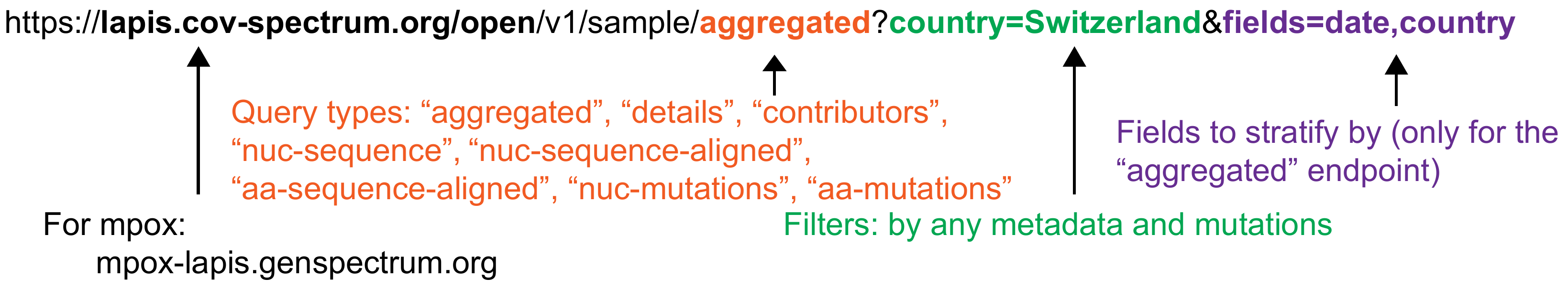}
    \caption{Components of a query link}
    \label{fig:query}
\end{figure}

The simplest way to use LAPIS is to encode a query in a URL prefixed with a particular LAPIS endpoint. Each LAPIS endpoint supports a different type of query and returns a different type of data (e.g., aggregated data, sequence data, mutations, etc.). Figure \ref{fig:query} illustrates a URL query structure. In the following sections, we explain the different parts of a query in more detail.

\subsubsection{Aggregation and stratification \label{sec:aggregation}}

LAPIS implements two types of endpoints: endpoints that provide aggregated data and endpoints that provide per-sample data. We describe the first type in this section and the second type in the next.

The \verb|aggregated| endpoint counts the number of samples that fulfill user-defined filters in a query. If the \verb|fields| parameter is not set, it returns the total number of samples. By setting \verb|fields|, we can stratify the data. E.g., \verb|/aggregated?fields=pangoLineage,country| will return the number of samples per Pango lineage and country. The fields parameter accepts all metadata and lineage-defining fields but not mutations or insertions.

To calculate the distribution of mutations and insertions, LAPIS offers the endpoints \verb|nuc-mutations|, \verb|aa-mutations|, \verb|nuc-insertions|, and \verb|aa-insertions|. They return the number of occurrences of mutations in a set of samples and their proportions. When calculating the proportions, the unknown or ambiguous bases are excluded. For example, if there are 10 sequences, 3 sequences have a mutation from \verb|A| to \verb|G| at position 5, 3 sequences have the reference base \verb|A|, and 4 sequences have an \verb|N| (i.e., unknown) at position 5, the proportion of the mutation \verb|A5G| is $\frac{3}{6} = 0.5$ (and not $\frac{3}{10} = 0.3$).

\subsubsection{Data download\label{sec:lookup}}

LAPIS can also be used to obtain non-aggregated data. The \verb|details| endpoint returns the metadata and supports an optional \verb|fields| parameter that can be used to limit the desired metadata fields. The \verb|nuc-sequence| and \verb|nuc-sequence-aligned| endpoints return the original and aligned nucleotide sequences, respectively. Finally, the \verb|aa-sequence-aligned/{gene}| endpoint (e.g., \verb|aa-sequence-aligned/S| for the SARS-CoV-2 Spike protein) returns the aligned amino acid sequences.

\subsubsection{Filters and advanced variant queries\label{sec:filters}}

By default (i.e., without specifying additional parameters), a query is evaluated on the whole set of sequences. To query a subset of sequences, a wide range of filters is available. It includes filtering by metadata, lineage names, and mutations. As shown in figure \ref{fig:query}, filters can be set by adding request parameters to the end of the URL. If multiple filters are set, the samples that fulfill all of them will be selected.

For ordinal data like dates, there are two available filters: one with a \verb|From|-suffix for the lower bound and one with a \verb|To|-suffix for the upper bound. E.g., \lstinline[breaklines=true]|dateFrom=2023-01-01&dateTo=2023-01-31| will filter for samples from January 2023.

LAPIS additionally supports two different ways to specify a variant. The simple approach is similar to the metadata filters and can be used to filter samples that fulfill all of a list of conditions. Possible parameters for the SARS-CoV-2 instance include \verb|pangoLineage|, \verb|aaMutations|, \verb|nucMutations|, \verb|aaInsertions|, etc. For the mutations/insertions, it is possible to use a comma-separated list. An example of a simple variant filter would be \verb|pangoLineage=XBB.1*&aaMutations=S:E484R,S:K417T|. The \verb|*| behind XBB.1 means all sub-lineages of the Pango lineage XBB.1 will also be included in the query. Insertion queries may contain wildcards, for instance, \verb|ins:1000:AAT?|. This filters for all sequences with an insertion that starts with \verb|AAT| between positions 1000 and 1001.

\begin{figure}
    \centering
    \includegraphics[width=\textwidth]{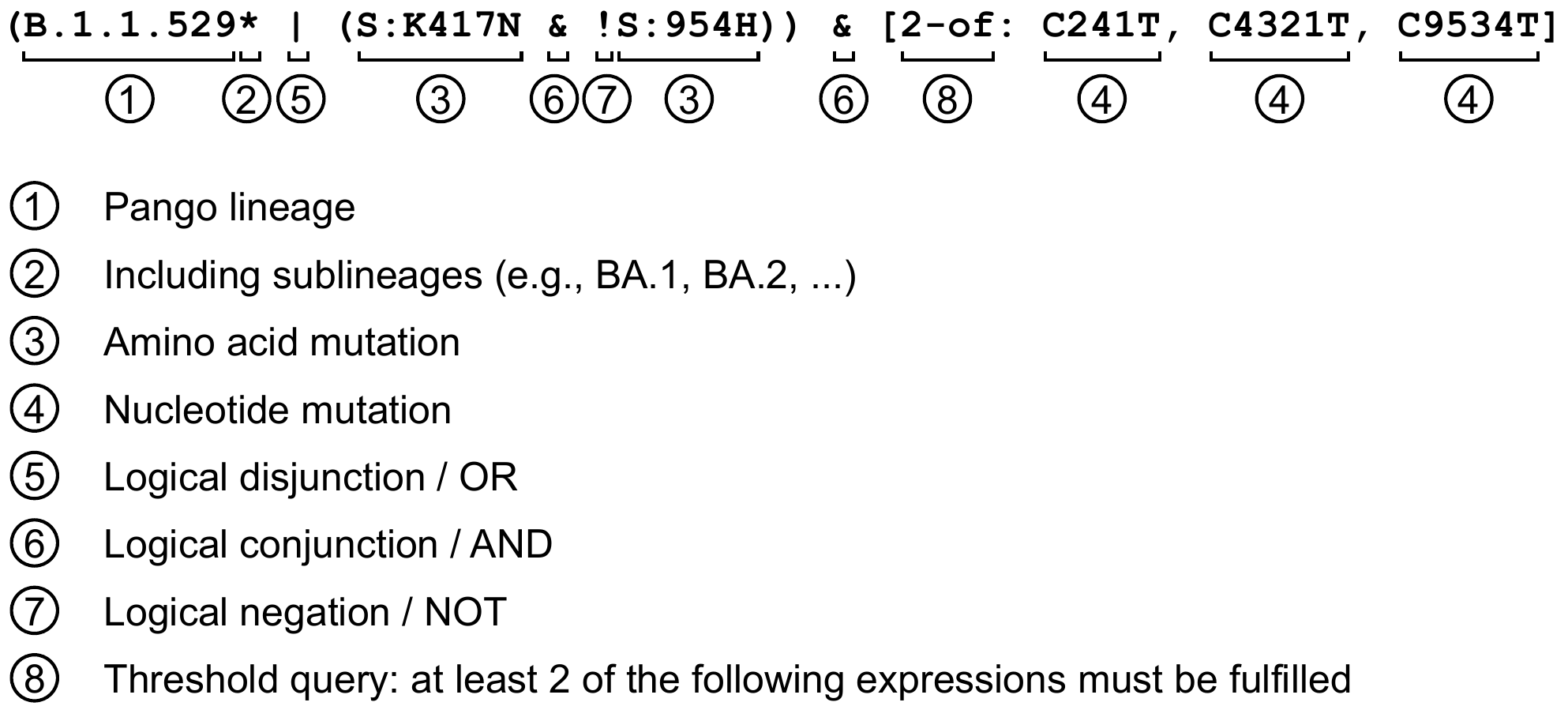}
    \caption{Components of an advanced variant query}
    \label{fig:advanced-query}
\end{figure}

The second approach is using advanced variant queries. Advanced variant queries support more than the conjunction of a list of conditions -- they also allow Boolean logic and threshold queries. One example is shown in figure \ref{fig:advanced-query}. Examples of real-world, user-defined advanced variant queries can be found in the CoV-Spectrum Collections\footnote{\href{https://cov-spectrum.org/collections}{https://cov-spectrum.org/collections}} where users can define and monitor sets of variants specified by advanced variant queries. In particular, the threshold queries have proven highly valuable. For example, they have been recently used to group sequences that share the same number of mutations in the receptor binding domain (RBD) \cite{Callaway2022-variant-soup}.

\subsubsection{``Maybe'' queries}\label{sec:maybe_queries}

Advanced variant queries further support ``maybe'' queries. These queries find sequences that \emph{might} have a certain mutation instead of definitely having a certain mutation. By default, when filtering for a mutation, LAPIS returns sequences for which the mutation is confirmed. E.g., the query \verb|A5G| selects sequences with a \verb|G| at position 5. This is a conservative way of filtering. In practice, we don't know the values at every position of every sequence: for some sequences, we might have a \verb|N| (=unknown/everything is possible) or another ambiguity code that includes \verb|G| such as \verb|K| (=\verb|G| or \verb|T|) at position 5. For those samples, it is possible that, in reality, they do have the mutation \verb|A5G|. This implies that the \verb|aggregated| endpoint usually\footnote{There is an exception for queries that contain a negation.} provides the lower-bound number of samples when we filter for mutations. ``maybe'' queries allow us to obtain the corresponding upper bound. For the \verb|A5G| example, sequences with a \verb|N|, \verb|X|, \verb|R|, \verb|S|, \verb|V|, \verb|D| and \verb|B| at position 5 will also be included. Maybe queries are part of the advanced variant queries. For example, we can query \verb|maybe(5G) & maybe(6T)|. In fact, we can write arbitrary variant query expressions in a \verb|maybe()| clause. Equivalent to the previous example, we can write \verb|maybe(5G & 6T)|. A more complex example would be \verb/maybe((S:10K & !S:11H) &  [2-of: 100A, 101T, 102G])/.

While the previous examples appear simple and intuitive, it is not always straightforward to determine the semantics of a maybe query. Let us consider the nucleotide sequence \verb|ATGCNT|. It has one unknown at position 5. The sequence would neither match the query \verb|5A| nor \verb|5C| but it would match \verb|maybe(5A)| and it would also match \verb|maybe(5C)|. What's about \verb|maybe(5A) & maybe(5C)|? From a Boolean logic perspective, if we consider \verb|maybe(5A)| and \verb|maybe(5C)| to be true, then their conjunction must be true as well. On the other hand, a sequence cannot have two different bases at the same position; thus, shouldn't \verb|maybe(5A) & maybe(5C)| be a contradiction and unconditionally false? LAPIS would evaluate \verb|maybe(5A) & maybe(5C)| for the aforementioned sequence to be true. The main reason we decided on this semantic is that it is possible to evaluate it efficiently\footnote{Although the alternative might appear more intuitive in some cases, we believe that it cannot be efficiently computed. We have not proven it formally but we think that the complexity of the alternative semantics grows exponentially with the length of the query.}.

\subsection{Performance\label{sec:speed}}

LAPIS is computationally efficient. It has proven capable of reliably processing millions of requests per day with most response times within a few hundred milliseconds as the backend to our CoV-Spectrum dashboard. 

\begin{figure}
    \centering
    \includegraphics[width=\textwidth]{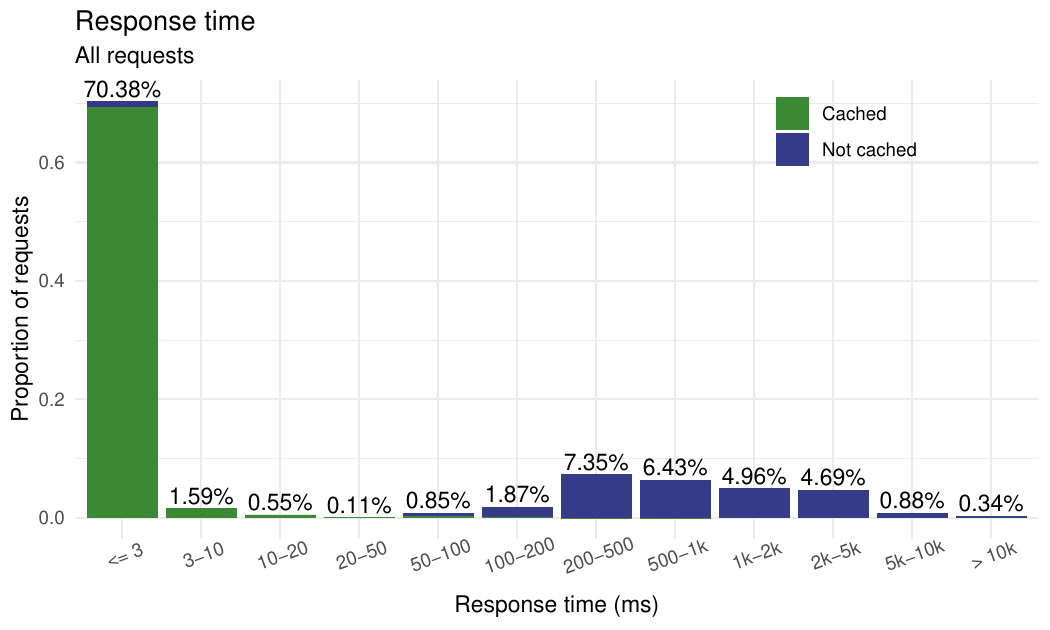}
    \caption{Proportion of requests for the different response time bins, stratified by cache status}
    \label{fig:response-time-all}
\end{figure}

We currently run the LAPIS instance for SARS-CoV-2 data from GISAID on an \verb|AWS r5.8xlarge| server (256 GB RAM, 32 vCPUs)\footnote{\href{https://aws.amazon.com/ec2/instance-types/}{https://aws.amazon.com/ec2/instance-types/}}. Between \update{25 January}\footnote{We only started logging the response times in the afternoon of 24 January 2023.} and \update{4 February 2023}, it processed \update{over 20 million} requests with a mean response time of \update{411~ms} and a median response time of \update{1~ms} (Table \ref{tab:endpoints}). This low median response time was possible because \update{72\%} of all responses had been cached (section \ref{sec:caching}), which greatly reduces response time (Figure \ref{fig:response-time-all}). Altogether, \update{83\%} of requests to the SARS-CoV-2 instance of LAPIS were processed within 500~ms.

\begin{figure}
    \centering
    \includegraphics[width=\textwidth]{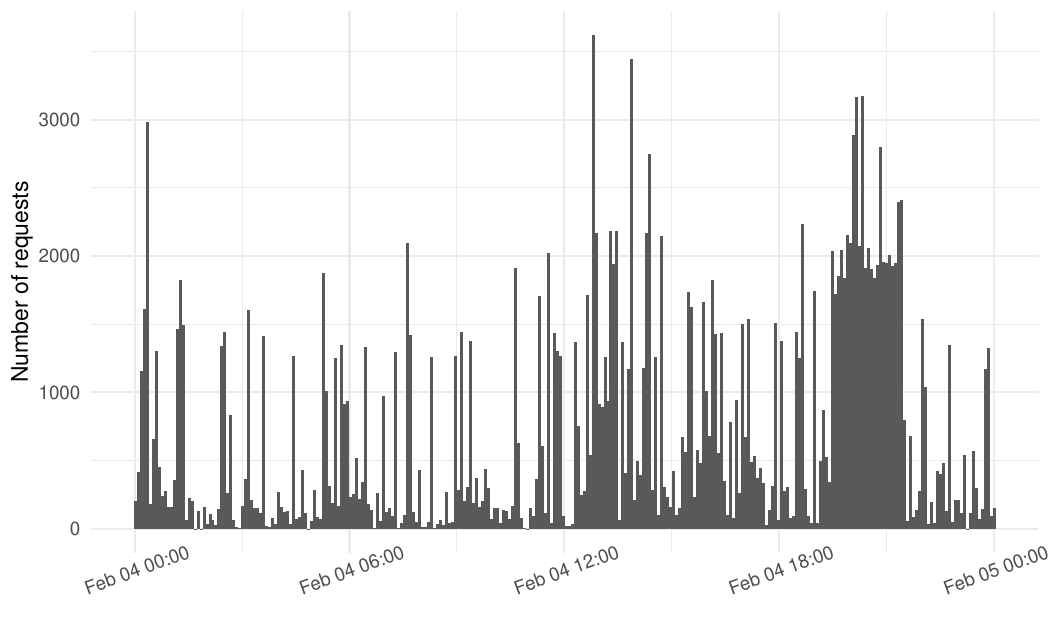}
    \caption{Number of requests within a day. Each bar represents one minute. In total, there were 208249 requests.}
    \label{fig:requests-over-time}
\end{figure}

LAPIS often has to process many requests in parallel. It is quite common to have very few requests in one minute and over a thousand in the next (Figure \ref{fig:requests-over-time}). The CoV-Spectrum collections are a major reason for that. In the user-defined collections, users can see information about many variants simultaneously. When a collection page is opened, the web application sends one request per variant to the server at the same time, and some collections (e.g., collection 24\footnote{\href{https://cov-spectrum.org/collections/24}{https://cov-spectrum.org/collections/24}}) have hundreds of variants. When we consider only requests that were executed when the server had less than 100 parallel requests (that is the case for \update{79\%} of the requests), \update{97\%} of the requests were processed within 500~ms.

In summary, the computational efficiency of LAPIS makes it suitable as a back-end for other tools and websites, including responsive and interactive dashboards and workflows. LAPIS achieves computational efficiency through a newly-developed data processing engine (see section \ref{sec:query_engine}) that is optimized for genomic data. It can perform common operations like searching for nucleotide mutations and amino acid changes in millions of sequences and hundreds of gigabytes of data within tens to hundreds of milliseconds.

\begin{table}[]
\caption{Empirical data on the usage and performance of the endpoints}
\small
\begin{tabular}{|l|l|l|l|l|}
\hline
Endpoint       & Number of requests    & Cache hit & Response time (mean/median, in ms)  \\ \hline
aa-insertions  & 34,212    (0.17\%)     & 14.19\%            & 654 / 254                \\ \hline
aa-mutations   & 101,346   (0.49\%)     & 20.86\%            & 1050 / 267                \\ \hline
aggregated     & 20,302,815 (98.85\%)    & 72.66\%            & 404 / 1                  \\ \hline
nuc-insertions & 34,221    (0.17\%)     & 14.20\%            & 639 / 252                \\ \hline
nuc-mutations  & 66,731    (0.32\%)     & 24.32\%            & 1203 / 266                \\ \hline
\end{tabular}
\label{tab:endpoints}
\end{table}

\begin{figure}
    \centering
    \includegraphics[width=\textwidth]{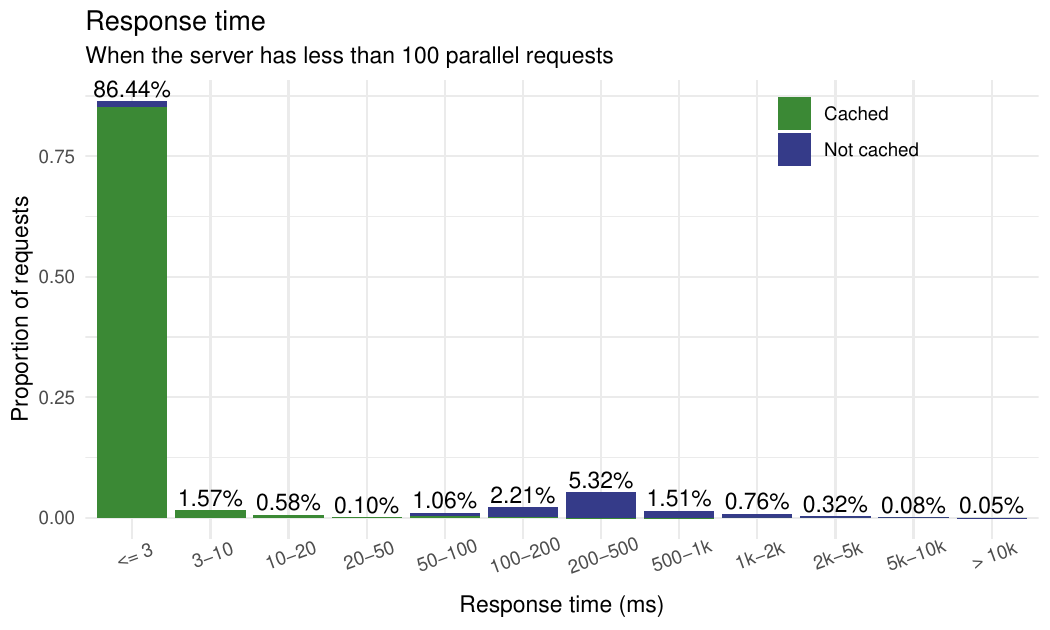}
    \caption{Same as figure \ref{fig:response-time-all} but only for requests executed when the server has less than 100 parallel requests}
    \label{fig:response-time-low-overlaps}
\end{figure}

\section{Discussion}\label{sec:discussion}

The unique filtering, aggregation, and download functionalities supported by LAPIS, coupled with high computational efficiency, make LAPIS a key resource for the real-time analysis of genomic sequencing data from ongoing outbreaks. LAPIS is currently available for all openly accessible SARS-CoV-2\footnote{\href{https://lapis.cov-spectrum.org/open}{https://lapis.cov-spectrum.org/open}} and mpox\footnote{\href{https://mpox-lapis.genspectrum.org}{https://mpox-lapis.genspectrum.org}} sequencing data on GenBank \cite{Benson2017}. We also maintain a private SARS-CoV-2 instance with sequencing data from GISAID \cite{Elbe2017-gisaid}, which serves as the backend for our CoV-Spectrum dashboard\footnote{\href{https://cov-spectrum.org}{https://cov-spectrum.org}}.

LAPIS' SARS-CoV-2 instances highlight the value of this approach as dataset size grows. As of January 2023, more than 14,500,000 SARS-CoV-2 sequences are available on GISAID, reaching a size of over 400 GB. LAPIS is capable of querying this entire dataset efficiently, supporting an interactive user experience on our CoV-Spectrum dashboard. CoV-Spectrum mainly presents aggregated data: it visualizes temporal, geographic, and mutational distributions of variants through a large variety of charts, tables, and maps. It solely uses LAPIS for retrieving genomic data and thanks to the flexibility of LAPIS, it was possible to develop new features in CoV-Spectrum without the need of extending or adapting LAPIS.

With LAPIS' mpox instance, we demonstrated the adaptability of the API approach. At the start of the mpox outbreak in 2022, within a few days of the release of the first sequence, we set up a LAPIS instance to support rapid sharing and easy access to open genomic data. It was accompanied by the MpoxSpectrum dashboard\footnote{\href{https://mpox.genspectrum.org}{https://mpox.genspectrum.org}} which, in addition to providing overview plots, enabled users to look up samples, download pre-processed metadata and aligned sequences, and open them in the Nextclade tool. To use the Nextclade integration feature, users can select sequences of interest on the MpoxSpectrum dashboard and Nextclade will download the sequences from LAPIS for quality analysis. Further, just four hours after we publicized LAPIS for mpox on Twitter, Taxonium announced the launch of a mpox service using LAPIS data as data source \cite{Sanderson_TaxoniumTweet2022, Sanderson_2022}.

%Discussion/Future work
%Future of LAPIS
These successes highlight that LAPIS fills a necessary role in addressing common challenges for accessing and analyzing genomic sequencing data. As demonstrated with mpox, LAPIS is easily extendable to other organisms. While supporting a new pathogen currently requires changes to the code base, we are actively working to generalize the LAPIS code to enable users to deploy instances with their own data and for other pathogens, possibly containing additional private metadata, via a configuration file. This will allow independent groups to run LAPIS instances for different use cases, akin to how Nextstrain publishes phylogenetic analyses for a limited number of pathogens but also provides the same analysis tools as an open-source resource for researchers to set up their own analyses. We hope to increase the incentive for data sharing in the public domain with this open-source philosophy: with the support of the API, researchers can directly analyze their own shared data within the global genomic context.

%open tasks
Going forward, we see great potential for database platforms such as GenBank to directly integrate APIs with functionalities like LAPIS's into their framework. This avoids the necessity of hosting data in a second database and allows researchers to benefit from functionality provided by an API such as LAPIS for many different organisms. 
On the research side, this requires developing techniques for efficiently querying even larger genomic data sets. The current implementation of LAPIS is capable of supporting up to around 20 to 30 million sequences of length 30kBp. We are working on better algorithms to push this boundary.

\section{Conclusions}

In summary, we introduce an in-memory database engine for genomic sequencing data which can be accessed through an API. This framework facilitates the analysis of millions of sequences in real time, meaning users can interactively query and filter sequencing data. In particular, our framework supports the analysis of open genomic sequencing data and enables researchers and authorities to rapidly analyze the evolution and epidemiology of pathogens for evidence-based public health response.

\section{Methods}

\subsection{Data pre-processing}\label{sec:preprocessing}

\begin{figure}[t]
    \centering
    \includegraphics[width=\textwidth]{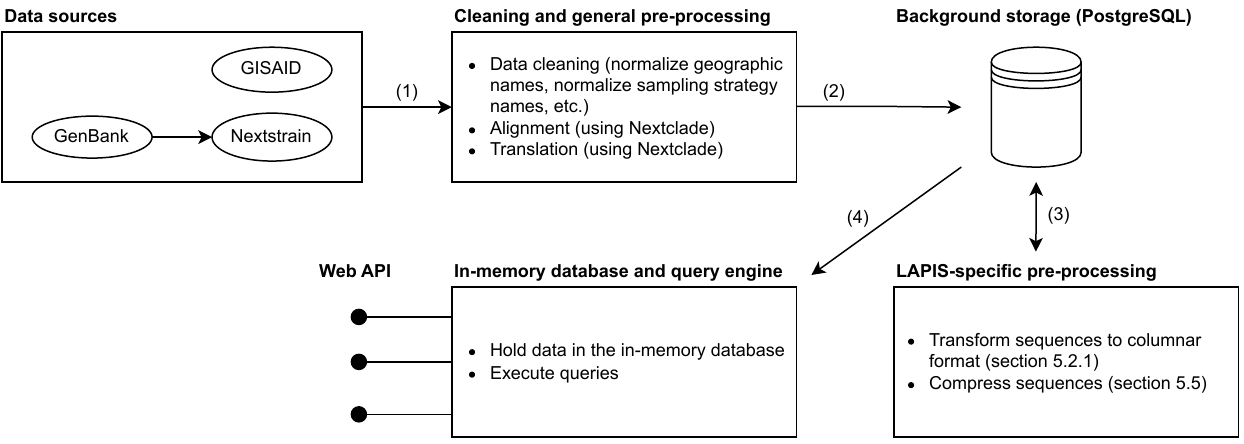}
    \caption{Data pre-processing workflow}
    \label{fig:overview}
\end{figure}

For the three LAPIS instances we currently maintain, we download the raw data from GISAID (SARS-CoV-2) or Nextstrain which retrieved it from GenBank (SARS-CoV-2 and mpox). The raw data contain the genomic (consensus) sequences and corresponding metadata. We pre-process the data in two steps. During the first step, we clean up the metadata, align the sequences to a reference genome, and translate the nucleotide sequences to protein amino acid sequences. For the alignment and translation, we use Nextclade \cite{Aksamentov2021} but other tools are equally applicable. The first step is not specific to LAPIS and can be replaced by alternative pipelines that produce an alignment and protein sequences. During the second step, we perform LAPIS-specific transformations and generate compressed columnar sequences (section \ref{sec:columnar_storage} and \ref{sec:compression}). The pre-processed data are loaded into the in-memory database (section \ref{sec:query_engine}) and exposed through a REST API. Figure \ref{fig:overview} illustrates the workflow.

We store the pre-processed data -- both after the first and after the second step -- in a PostgreSQL database. Hereby, the PostgreSQL database only serves as a background storage and can be easily replaced by the file system or a different database system. It is not crucial to the performance of LAPIS outside of the pre-processing pipeline because the in-memory database is used for evaluating the queries.

\subsection{Data query engine}\label{sec:query_engine}

We developed a novel data query engine for our public web API that is tailored to support real-time, interactive genomic surveillance and genomic epidemiology. Specifically, it is designed to support high numbers of requests and fast query processing of genomic sequencing data. Our internal SARS-CoV-2 LAPIS instance based on GISAID data currently receives hundreds of thousands of requests per day, mostly from users of CoV-Spectrum. At the same time, it must support interactive and exploratory analyses where the user is able to switch quickly between different variants, countries, and time periods by responding to most requests within tens to hundreds of milliseconds. Existing database systems are not sufficient for this task.

\subsubsection{Column-wise storage}\label{sec:columnar_storage}

\begin{figure}[t]
    \centering
    \includegraphics[width=\textwidth]{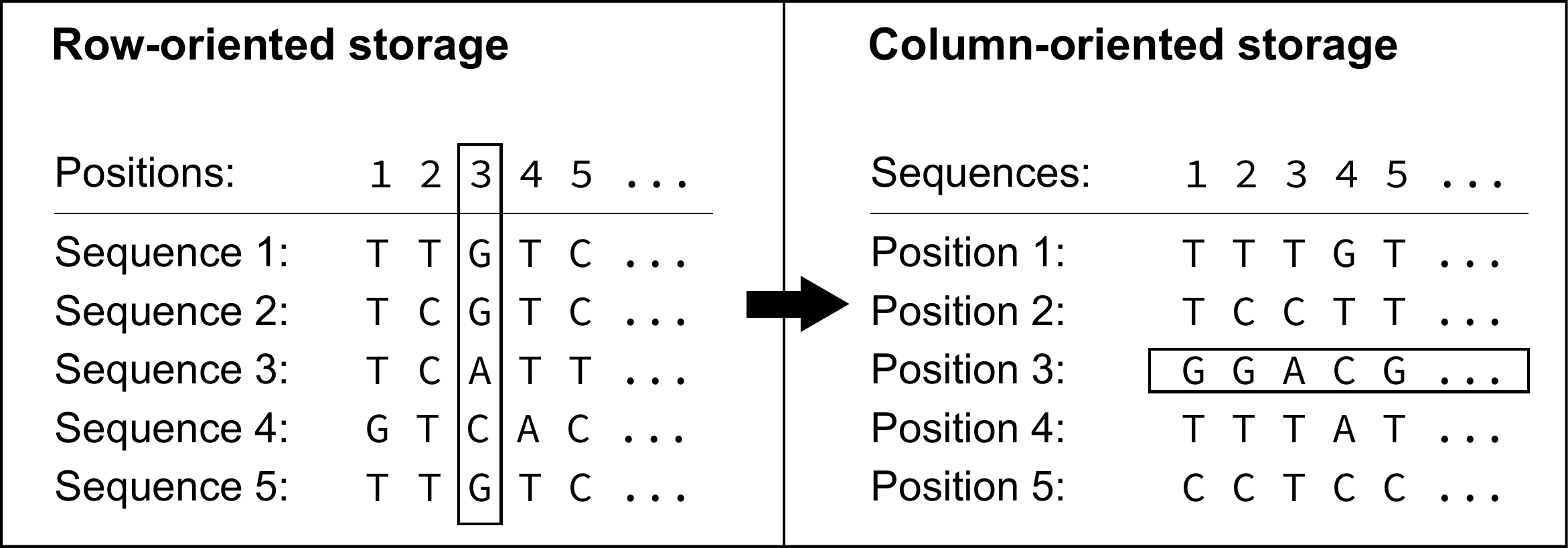}
    \caption{Transformation of sequences to the columnar format: The row-oriented storage maintains one string per sequence; in contrast, the column-oriented storage keeps one string per position.}
    \label{fig:columnar-format}
\end{figure}

Our approach is based on techniques developed for column-oriented database systems \cite{Abadi_2012}. In the pre-processing step, we transform the sequencing data into a columnar format. For each position in the aligned nucleotide sequence or in the aligned amino acid sequence, we construct a string with the characters of all sequences at that position (figure \ref{fig:columnar-format}). The $i$-th character in the new, columnar sequence corresponds to the sequence with the ID $i$. To find sequences with a mutation at a given position, we then only need to read a single string and not filter through each sequence. The columnar sequences are easy to compress (section \ref{sec:compression}), and by compressing them, we can cache them in memory and eliminate any disk and round-trip time to the database.

\subsubsection{Filter insertions}

The column store as described in the previous subsection can only store the aligned sequences. It has one column for each base of the reference genome but it cannot store insertions which are parts of a sequence that cannot be directly linked to the positions of the reference. To filter for insertions, LAPIS uses a dedicated insertion store which maintains for each position of the reference genome a mapping of inserted values to sequences with the insertion. E.g., a mapping of \verb|AATGGC| at position 1000 to \verb|{sequence1, sequence2, sequence3}| means that there are three sequences that have the insertion \verb|AATGGC| between position 1000 and 1001.

To evaluate a query such as \verb|ins:1000:AAT?| as described in section \ref{sec:filters}, LAPIS looks up the insertions at position 1000 and matches them against the requested pattern. For the SARS-CoV-2 data, this approach works well because insertions are rare, short, and not very diverse. For genomic data with many long and diverse insertions, this method is not very efficient.

\subsubsection{Sequence downloads}

To download whole sequences, LAPIS first filters the sequences with the in-memory query engine, fetches compressed sequences (section \ref{sec:compression}) from the background storage (section \ref{sec:preprocessing}), and decompresses them. If a large set of sequences should be downloaded, it fetches the sequences in small batches and streams them to the user to ensure a low memory footprint.

\subsubsection{Discussion}

The data engine was first deployed when there were around one million genomic sequences for SARS-CoV-2, and it still performs well for 15 million sequences today. It is a significant improvement to using common relational database systems which are not optimized for genomic sequencing data. The current algorithm is simple and easy to implement. However, it is also rather naive and not using state-of-the-art database engineering techniques. We are working on an improved version with reduced response times and higher throughput and look forward to sharing our results in the near future.

\subsection{Data versions}\label{sec:data_versions}

To allow the user to use consistent data, each response of LAPIS contains the version of the data. The user can then check if the data versions of multiple requests are the same, and reload if that is not the case. The data version is provided in the HTTP response header \verb|LAPIS-Data-Version|. For JSON responses, the data version is further given in the \verb|dataVersion| field.

For example, this is relevant to compute the proportion of a variant in the sequencing data. For the calculation, we would fetch the number of sequences of the variant and the total number of sequences; that means that two API calls are required. In this case, data could be updated between the two calls which would lead to wrong results because the nominator and denominator to calculate the proportion are incompatible. Comparing the data versions of the two requests would prevent an error.

\subsection{Caching}\label{sec:caching}

To minimize the response time for common requests, LAPIS caches the results of previously evaluated queries in a Redis database. Caching is usually a difficult task due to the complexity to determine when a cache entry is stale. In the case of LAPIS, however, we have the advantage that we do not have a continuous stream of small data changes but perform rare (e.g., once a day) but big updates. This allows us to distinguish different versions of the data (section \ref{sec:data_versions}).

Each cached result is linked to a data version. If the user defines a data version in a request, and the result generated from the data of the specified version is in the cache, it can be returned immediately. If the user does not define a data version, LAPIS will check if the result for the most recent data version is cached. Figure \ref{fig:response-time-all} and \ref{fig:response-time-low-overlaps} and table \ref{tab:endpoints} show the proportions of cache hits.

\subsection{Compression}\label{sec:compression}

We compress the genome sequences before inserting them into the database. We use Zstd (level 3) \cite{Collet2021} which gives us a good balance between compression ratio and speed. For the compression of the whole nucleotide and amino acid sequences, we use the respective reference sequence as the pre-defined dictionary to improve the compression ratio. For the columnar sequences, a pre-defined dictionary is not needed for a good compression ratio as it is intrinsically easy to compress. We achieve a compression ratio of 94\% for the unaligned sequences, 99.3\% for the aligned sequences, and 96\% for the sequences stored in the column-oriented format.

\section{Declarations}

\subsection{Ethics approval and consent to participate}

Not applicable

\subsection{Availability of data and materials}

The code is released under the GPL-3.0 license at \href{https://github.com/GenSpectrum/LAPIS}{https://github.com/GenSpectrum/LAPIS}.

The current study did not generate new genomic sequencing data. Part of the analyzed data is publicly available in the INSDC (GenBank/ENA/DDBJ) repositories (https://www.insdc.org/). The remaining part of the analyzed SARS-CoV-2 data is available in GISAID (https://gisaid.org/) to which everyone with a GISAID account has access.

\subsection{Competing interests}

The authors declare that they have no competing interests.

\subsection{Consent for publication}

Not applicable.

\subsection{Funding}
The authors acknowledge funding through the Swiss National Science Foundation (31CA30$\_$196267), the Federal Office of Public Health, Switzerland (Dienstleistungsvertrag ``Analysen zur Überwachung von SARS-CoV-2: Varianten-Monitoring und now-casting Hospitalisationen mit Zusatzoption Ad-hoc Analysen''), and ETH Z\"{u}rich.

\subsection{Authors' contributions}

\begin{itemize}[leftmargin=*]
    \setlength\itemsep{0em}
    \item CC: Conceptualization, Data curation, Formal Analysis, Investigation, Methodology, Software, Visualization, Supervision, Writing – original draft
    \item AT: Methodology, Software, Writing – review \& editing
    \item FE: Software, Writing – review \& editing
    \item JK: Software, Writing – review \& editing
    \item CR: Conceptualization, Writing – review \& editing
    \item TS: Conceptualization, Funding acquisition, Resources, Supervision, Writing – original draft
\end{itemize}

\section{Acknowledgements}
We acknowledge all labs for the timely submission of their SARS-CoV-2 and mpox  sequencing data to the ISNDC (GenBank, ENA, and DDBJ) and GISAID. We acknowledge the Nextstrain team for pre-processing the GenBank data. These efforts are core for LAPIS to share sequencing data from outbreaks as they are generated. We thank \orcidaffil{0000-0003-1008-8918} Sarah Nadeau for valuable discussions and feedback on the manuscript. We thank \orcidaffil{0000-0002-5396-1193} François Bienvenu for valuable discussions on the ``maybe'' queries.

\printbibliography

@Article{Chen2021-b117,
  author   = {Chen, Chaoran and Nadeau, Sarah Ann and Topolsky, Ivan and Manceau, Marc and Huisman, Jana S. and Jablonski, Kim Philipp and Fuhrmann, Lara and Dreifuss, David and Jahn, Katharina and Beckmann, Christiane and Redondo, Maurice and Noppen, Christoph and Risch, Lorenz and Risch, Martin and Wohlwend, Nadia and Kas, Sinem and Bodmer, Thomas and Roloff, Tim and Stange, Madlen and Egli, Adrian and Eckerle, Isabella and Kaiser, Laurent and Denes, Rebecca and Feldkamp, Mirjam and Nissen, Ina and Santacroce, Natascha and Burcklen, Elodie and Aquino, Catharine and de Gouvea, Andreia Cabral and Moccia, Maria Domenica and Grüter, Simon and Sykes, Timothy and Opitz, Lennart and White, Griffin and Neff, Laura and Popovic, Doris and Patrignani, Andrea and Tracy, Jay and Schlapbach, Ralph and Dermitzakis, Emmanouil T. and Harshman, Keith and Xenarios, Ioannis and Pegeot, Henri and Cerutti, Lorenzo and Penet, Deborah and Blin, Anthony and Elies, Melyssa and Althaus, Christian L. and Beisel, Christian and Beerenwinkel, Niko and Ackermann, Martin and Stadler, Tanja},
  journal  = {Epidemics},
  title    = {Quantification of the spread of {SARS}-{CoV}-2 variant {B}.1.1.7 in {Switzerland}},
  year     = {2021},
  issn     = {1755-4365},
  month    = dec,
  pages    = {100480},
  volume   = {37},
  doi      = {10.1016/j.epidem.2021.100480},
  language = {en}
}

@Article{Black2020,
  author     = {Allison Black and Duncan R. MacCannell and Thomas R. Sibley and Trevor Bedford},
  journal    = {Nature Medicine},
  title      = {Ten recommendations for supporting open pathogen genomic analysis in public health},
  year       = {2020},
  month      = {jun},
  number     = {6},
  pages      = {832--841},
  volume     = {26},
  doi        = {10.1038/s41591-020-0935-z},
  publisher  = {Springer Science and Business Media {LLC}},
  readstatus = {read},
}

@Article{Benson2017,
  author    = {Dennis A Benson and Mark Cavanaugh and Karen Clark and Ilene Karsch-Mizrachi and James Ostell and Kim D Pruitt and Eric W Sayers},
  journal   = {Nucleic Acids Research},
  title     = {{GenBank}},
  year      = {2017},
  month     = {nov},
  number    = {D1},
  pages     = {D41--D47},
  volume    = {46},
  doi       = {10.1093/nar/gkx1094},
  publisher = {Oxford University Press ({OUP})},
}

@Article{Tegally2022,
  author    = {Houriiyah Tegally and Monika Moir and Josie Everatt and Marta Giovanetti and Cathrine Scheepers and Eduan Wilkinson and Kathleen Subramoney and Sikhulile Moyo and Daniel G. Amoako and Cheryl Baxter and Christian L. Althaus and Ugochukwu J. Anyaneji and Dikeledi Kekana and Raquel Viana and Jennifer Giandhari and Richard J. Lessells and Tongai Maponga and Dorcas Maruapula and Wonderful Choga and Mogomotsi Matshaba and Simnikiwe Mayaphi and Nokuzola Mbhele and Mpaphi B. Mbulawa and Nokukhanya Msomi and Yeshnee Naidoo and Sureshnee Pillay and Tomasz Janusz Sanko and James E. San and Lesley Scott and Lavanya Singh and Nonkululeko A. Magini and Pamela Smith-Lawrence and Wendy Stevens and Graeme Dor and Derek Tshiabuila and Nicole Wolter and Wolfgang Preiser and Florette K. Treurnicht and Marietjie Venter and Michaela Davids and Georginah Chiloane and Adriano Mendes and Caitlyn McIntyre and Aine O'Toole and Christopher Ruis and Thomas P. Peacock and Cornelius Roemer and Carolyn Williamson and Oliver G. Pybus and Jinal Bhiman and Allison Glass and Darren P. Martin and Andrew Rambaut and Simani Gaseitsiwe and Anne von Gottberg and Tulio de Oliveira and},
  title     = {Continued Emergence and Evolution of Omicron in South Africa: New {BA}.4 and {BA}.5 lineages},
  year      = {2022},
  month     = {may},
  doi       = {10.1101/2022.05.01.22274406},
  publisher = {Cold Spring Harbor Laboratory},
}

@Article{Viana2022,
  author    = {Raquel Viana and Sikhulile Moyo and Daniel G. Amoako and Houriiyah Tegally and Cathrine Scheepers and Christian L. Althaus and Ugochukwu J. Anyaneji and Phillip A. Bester and Maciej F. Boni and Mohammed Chand and Wonderful T. Choga and Rachel Colquhoun and Michaela Davids and Koen Deforche and Deelan Doolabh and Louis du Plessis and Susan Engelbrecht and Josie Everatt and Jennifer Giandhari and Marta Giovanetti and Diana Hardie and Verity Hill and Nei-Yuan Hsiao and Arash Iranzadeh and Arshad Ismail and Charity Joseph and Rageema Joseph and Legodile Koopile and Sergei L. Kosakovsky Pond and Moritz U. G. Kraemer and Lesego Kuate-Lere and Oluwakemi Laguda-Akingba and Onalethatha Lesetedi-Mafoko and Richard J. Lessells and Shahin Lockman and Alexander G. Lucaci and Arisha Maharaj and Boitshoko Mahlangu and Tongai Maponga and Kamela Mahlakwane and Zinhle Makatini and Gert Marais and Dorcas Maruapula and Kereng Masupu and Mogomotsi Matshaba and Simnikiwe Mayaphi and Nokuzola Mbhele and Mpaphi B. Mbulawa and Adriano Mendes and Koleka Mlisana and Anele Mnguni and Thabo Mohale and Monika Moir and Kgomotso Moruisi and Mosepele Mosepele and Gerald Motsatsi and Modisa S. Motswaledi and Thongbotho Mphoyakgosi and Nokukhanya Msomi and Peter N. Mwangi and Yeshnee Naidoo and Noxolo Ntuli and Martin Nyaga and Lucier Olubayo and Sureshnee Pillay and Botshelo Radibe and Yajna Ramphal and Upasana Ramphal and James E. San and Lesley Scott and Roger Shapiro and Lavanya Singh and Pamela Smith-Lawrence and Wendy Stevens and Amy Strydom and Kathleen Subramoney and Naume Tebeila and Derek Tshiabuila and Joseph Tsui and Stephanie van Wyk and Steven Weaver and Constantinos K. Wibmer and Eduan Wilkinson and Nicole Wolter and Alexander E. Zarebski and Boitumelo Zuze and Dominique Goedhals and Wolfgang Preiser and Florette Treurnicht and Marietje Venter and Carolyn Williamson and Oliver G. Pybus and Jinal Bhiman and Allison Glass and Darren P. Martin and Andrew Rambaut and Simani Gaseitsiwe and Anne von Gottberg and Tulio de Oliveira},
  journal   = {Nature},
  title     = {Rapid epidemic expansion of the {SARS}-{CoV}-2 Omicron variant in southern Africa},
  year      = {2022},
  month     = {jan},
  number    = {7902},
  pages     = {679--686},
  volume    = {603},
  doi       = {10.1038/s41586-022-04411-y},
  publisher = {Springer Science and Business Media {LLC}},
}

@Article{Li_2021,
  author    = {Juan Li and Shengjie Lai and George F. Gao and Weifeng Shi},
  journal   = {Nature},
  title     = {The emergence, genomic diversity and global spread of {SARS}-{CoV}-2},
  year      = {2021},
  month     = {dec},
  number    = {7889},
  pages     = {408--418},
  volume    = {600},
  doi       = {10.1038/s41586-021-04188-6},
  publisher = {Springer Science and Business Media {LLC}},
}

@Article{Chen2021-covspectrum,
  author    = {Chaoran Chen and Sarah Nadeau and Michael Yared and Philippe Voinov and Ning Xie and Cornelius Roemer and Tanja Stadler},
  journal   = {Bioinformatics},
  title     = {{CoV}-Spectrum: analysis of globally shared {SARS}-{CoV}-2 data to identify and characterize new variants},
  year      = {2021},
  month     = {dec},
  number    = {6},
  pages     = {1735--1737},
  volume    = {38},
  doi       = {10.1093/bioinformatics/btab856},
  publisher = {Oxford University Press ({OUP})},
}

@Article{Hodcroft2021,
  author    = {Emma B. Hodcroft and Nicola De Maio and Rob Lanfear and Duncan R. MacCannell and Bui Quang Minh and Heiko A. Schmidt and Alexandros Stamatakis and Nick Goldman and Christophe Dessimoz},
  journal   = {Nature},
  title     = {Want to track pandemic variants faster? Fix the bioinformatics bottleneck},
  year      = {2021},
  month     = {mar},
  number    = {7848},
  pages     = {30--33},
  volume    = {591},
  doi       = {10.1038/d41586-021-00525-x},
  publisher = {Springer Science and Business Media {LLC}},
}

@Article{Abadi_2012,
  author    = {Daniel Abadi},
  journal   = {Foundations and Trends{\textregistered} in Databases},
  title     = {The Design and Implementation of Modern Column-Oriented Database Systems},
  year      = {2012},
  issn      = {1931-7883, 1931-7891},
  number    = {3},
  pages     = {197--280},
  volume    = {5},
  doi       = {10.1561/1900000024},
  language  = {en},
  publisher = {Now Publishers},
  urldate   = {2016-12-03},
}

@Article{Knyazev2022,
  author    = {Sergey Knyazev and Karishma Chhugani and Varuni Sarwal and Ram Ayyala and Harman Singh and Smruthi Karthikeyan and Dhrithi Deshpande and Pelin Icer Baykal and Zoia Comarova and Angela Lu and Yuri Porozov and Tetyana I. Vasylyeva and Joel O. Wertheim and Braden T. Tierney and Charles Y. Chiu and Ren Sun and Aiping Wu and Malak S. Abedalthagafi and Victoria M. Pak and Shivashankar H. Nagaraj and Adam L. Smith and Pavel Skums and Bogdan Pasaniuc and Andrey Komissarov and Christopher E. Mason and Eric Bortz and Philippe Lemey and Fyodor Kondrashov and Niko Beerenwinkel and Tommy Tsan-Yuk Lam and Nicholas C. Wu and Alex Zelikovsky and Rob Knight and Keith A. Crandall and Serghei Mangul},
  journal   = {Nature Methods},
  title     = {Unlocking capacities of genomics for the {COVID}-19 response and future pandemics},
  year      = {2022},
  month     = {apr},
  doi       = {10.1038/s41592-022-01444-z},
  publisher = {Springer Science and Business Media {LLC}},
}

@Article{Aksamentov2021,
  author     = {Ivan Aksamentov and Cornelius Roemer and Emma Hodcroft and Richard Neher},
  journal    = {Journal of Open Source Software},
  title      = {Nextclade: clade assignment, mutation calling and quality control for viral genomes},
  year       = {2021},
  month      = {nov},
  number     = {67},
  pages      = {3773},
  volume     = {6},
  doi        = {10.21105/joss.03773},
  publisher  = {The Open Journal},
  shorttitle = {Nextclade},
}

@Misc{Sanderson_TaxoniumTweet2022,
  author    = {Sanderson,Theo},
  month     = may,
  title     = {Tweet},
  year      = {2022},
  url       = {https://twitter.com/theosanderson/status/1528469460482502659},
}

@Article{Sanderson_2022,
article_type = {journal},
title = {Taxonium, a web-based tool for exploring large phylogenetic trees},
author = {Sanderson, Theo},
volume = 11,
year = 2022,
month = {nov},
pub_date = {2022-11-15},
pages = {e82392},
citation = {eLife 2022;11:e82392},
doi = {10.7554/eLife.82392},
url = {https://doi.org/10.7554/eLife.82392},
journal = {eLife},
issn = {2050-084X},
publisher = {eLife Sciences Publications, Ltd},
}

@Misc{Hodcroft_CoVariants2021,
  author = {Hodcroft,Emma B.},
  title  = {CoVariants: SARS-CoV-2 Mutations and Variants of Interest},
  year   = {2021},
  url    = {https://covariants.org/},
}

@Misc{cdc-tracker,
   author={{Centers for Disease Control and Prevention}},
   title={CDC Covid Data tracker},
   url={https://covid.cdc.gov/covid-data-tracker},
   year = {2023}
}

@Article{Gangavarapu-outbreakinfo,
  author    = {Karthik Gangavarapu and Alaa Abdel Latif and Julia L. Mullen and Manar Alkuzweny and Emory Hufbauer and Ginger Tsueng and Emily Haag and Mark Zeller and Christine M. Aceves and Karina Zaiets and Marco Cano and Xinghua Zhou and Zhongchao Qian and Rachel Sattler and Nathaniel L. Matteson and Joshua I. Levy and Raphael T. C. Lee and Lucas Freitas and Sebastian Maurer-Stroh and {GISAID Core and Curation Team} and Marc A. Suchard and Chunlei Wu and Andrew I. Su and Kristian G. Andersen and Laura D. Hughes},
  journal   = {Nature Methods},
  title     = {Outbreak.info genomic reports: scalable and dynamic surveillance of {SARS}-{CoV}-2 variants and mutations},
  year      = {2023},
  month     = {feb},
  doi       = {10.1038/s41592-023-01769-3},
  publisher = {Springer Science and Business Media {LLC}},
}

@Standard{Collet2021,
  organization = {Internet Engineering Task Force (IETF)},
  title        = {RFC 8878: Zstandard Compression and the 'application/zstd' Media Type},
  author       = {Y. Collet},
  url          = {https://datatracker.ietf.org/doc/html/rfc8878},
  year         = {2021},
}

@Article{Elbe2017-gisaid,
  author     = {Elbe, Stefan and Buckland-Merrett, Gemma},
  journal    = {Global Challenges},
  title      = {Data, disease and diplomacy: {GISAID}'s innovative contribution to global health},
  year       = {2017},
  issn       = {2056-6646},
  number     = {1},
  pages      = {33--46},
  volume     = {1},
  doi        = {10.1002/gch2.1018},
  language   = {en},
  shorttitle = {Data, disease and diplomacy},
  url        = {https://onlinelibrary.wiley.com/doi/abs/10.1002/gch2.1018},
  urldate    = {2022-04-04},
}

@Article{Callaway2022-variant-soup,
  author    = {Ewen Callaway},
  journal   = {Nature},
  title     = {{COVID} `variant soup' is making winter surges hard to predict},
  year      = {2022},
  month     = {oct},
  number    = {7935},
  pages     = {213--214},
  volume    = {611},
  doi       = {10.1038/d41586-022-03445-6},
  publisher = {Springer Science and Business Media {LLC}},
}

@Article{Chen2022-challenges,
  author    = {Chaoran Chen and Sarah Nadeau and Ivan Topolsky and Niko Beerenwinkel and Tanja Stadler},
  journal   = {Epidemics},
  title     = {Advancing genomic epidemiology by addressing the bioinformatics bottleneck: Challenges, design principles, and a Swiss example},
  year      = {2022},
  month     = {jun},
  pages     = {100576},
  volume    = {39},
  doi       = {10.1016/j.epidem.2022.100576},
  publisher = {Elsevier {BV}},
}
% \bibliographystyle{abbrv}
% \bibliography{main}

\end{document}